\documentclass[useAMS,usenatbib,usedcolumn]{mn2e}

\usepackage{graphicx}

\input{tmp.cls}

\def\cf{{cf.~}}
\def\eg{{e.g.,~}}
\def\ie{{i.e.,~}}
\def\lsim{\raise0.3ex\hbox{$<$}\kern-0.75em{\lower0.65ex\hbox{$\sim$}}} 
\def\gsim{\raise0.3ex\hbox{$>$}\kern-0.75em{\lower0.65ex\hbox{$\sim$}}} 
\def\sc1{\raise2.1ex\hbox{\tiny $r\!\!=\!\!4$}\kern-0.95em{\hbox{$=$}}}

\def\cm3{~{\rm cm^{-3}}}

\def\hinv{$h^{-1}$}

\def\hxr{hard X-ray~}
\def\ltsima{$\; \buildrel < \over \sim \;$}
\def\simlt{\lower.5ex\hbox{\ltsima}}
\def\gtsima{$\; \buildrel > \over \sim \;$}
\def\simgt{\lower.5ex\hbox{\gtsima}}

\def\sc{{\rm Science\ }}

\newcommand{\igm}{inter-galactic medium }

\newcommand{\gr}{$\gamma$-ray }
\newcommand{\epm}{e$^\pm$ }
\newcommand{\cre}{e$^-$ }
\newcommand{\pnd}{$\pi^0$-decay }

\newcommand{\ic}{IC }

\begin{document}
\title[]
  {Numerical Modeling of Gamma Radiation from Galaxy Clusters}
\author[F.\ Miniati]
  {Francesco~Miniati\thanks{fm@MPA-Garching.MPG.DE}\\
  Max-Planck-Institut f\"ur Astrophysik,
     Karl-Schwarzschild-Str. 1, 85740, Garching, Germany}
\date{\today}
\pubyear{2001} \volume{000} \pagerange{1} \onecolumn

\maketitle \label{firstpage}

\begin{abstract}
We investigate the spatial and spectral properties of 
non-thermal emission from clusters of galaxies at \gr energies.
We estimate the radiation flux between 10 keV and 10 TeV
due to inverse-Compton (IC) emission, \pnd and non-thermal 
bremsstrahlung (NTB) from cosmic-ray (CR) ions and electrons 
accelerated at cosmic shocks as well
as secondary \epm generated in inelastic p-p collisions.
We identify two main region of production of non-thermal radiation,
namely the core (also bright in thermal X-ray) 
and the outskirts region where accretion shocks occur.
We find that \ic emission from shock accelerated CR electrons
dominate the emission at the outer regions of galaxy clusters,
provided that at least a fraction of a percent of the shock 
ram pressure is converted into CR electrons.
A clear detection of this component and of its spatial 
distribution will allow us direct probing of cosmic accretion shocks.
In the cluster core, \gr emission above 100 MeV is dominated
by \pnd mechanism. At lower energies, \ic emission from 
secondary \epm takes over. However, \ic emission from 
shock accelerated electrons projected onto the cluster core
will not be negligible. We emphasize the importance of
separating the aforementioned emission components 
for a correct interpretation of the experimental data and
outline a strategy for that purpose. Failure in addressing this 
issue will produce unsound estimates of the intra-cluster
magnetic field strength and CR ion content.
According to our estimate future space borne and ground based 
\gr facilities should be able to measure the whole non-thermal 
spectrum both in the cluster core and at its outskirts.
The importance of such measurements in advancing our understanding
of non-thermal processes in the intra-cluster medium is discussed.

\end{abstract}


\section{Introduction}
\label{intro.se}

The structure and evolution of the large scale universe
is a fundamental tool of cosmological investigation.
Clusters of galaxies in particular, being the largest bound 
objects in the universe, have been studied extensively
in order to ``weight'' the cosmos and to probe its structure.
In the currently favored cold dark matter (CDM) 
hierarchical scenarios structure formation originates
from the collapse of primordial density perturbations
seeded at an inflationary epoch \citep{peebles93}.
In this depiction,
rich clusters of galaxies correspond to linear 
perturbations on scales of order $\sim 10$ Mpc comoving
and, therefore, their potential wells are expected to 
accommodate a representative fraction of the actual 
mass content of the universe \citep{wnef93}. 
By the same token, the environment there should be the end 
result of physical processes that occurred below or at that 
length scale. 
And, in fact, observational evidence tells us that
with respect to its pristine conditions the
intra-cluster medium (ICM) gas has been 
metal enriched by star formation processes, magnetized
in some fashion and shock heated.

Perhaps spurred by observational progress, in recent years
the subject of non-thermal process in, and their impact on, 
the ICM has attracted 
much attention. In this regard, \cite{voahbr96} pointed out 
that metal enrichment by galactic winds is likely accompanied 
by inter-galactic termination shocks where cosmic-ray (CR) 
acceleration might take place. In addition, \cite{ebkw97} studied
the case of CR protons escaping from of radio jets into the ICM
and \cite{krdulico01} 
that of black hole magnetic energy injection into the \igm (IGM).
The issue of CR ions produced at cosmic shocks \citep{minetal00}
and their possible dynamical role was addressed by 
\cite{mrkj01} by means of a cosmological simulation of
structure formation which included direct treatment of 
acceleration, transport and energy losses of a CR component.
They found that if shock acceleration takes place with some
efficiency, CR ions, due to their
long lifetime against energy losses and to their efficient 
confinement by magnetic irregularities \citep{voahbr96,bbp97}, 
may accumulate inside forming structures and store a 
significant fraction of the total pressure
there \citep{mint00,mrkj01}.

In order to assess the level of non-thermal activity in groups 
and clusters direct observations of the associated non-thermal 
emission is required. In particular, the integrated \gr photon 
flux above 100 MeV produced in the decay of neutral $\pi$-mesons 
generated in p-p inelastic collisions, allows direct
determination of the CR ion content and the pressure support that 
they provide. For the sources of CRs mentioned in the previous 
paragraph, the authors that investigated them estimated 
a \gr flux above 100 MeV that should be measurable by
future targeted observations 
\citep{voahbr96,bbp97,ebkw97,blasi99,atvo00,mrkj01}.
However, in a recent paper \citet{min02b} computed the
inverse Compton (IC) \gr emission from shock accelerated CR 
electrons scattering off Cosmic Microwave Background (CMB)
photons and for typical clusters of
galaxies found it comparable to the \gr
emission associated to $\pi^0$-decay. 
The importance of the former process was recently pointed out by
\citet{lowa00} in the context of the unexplained cosmic \gr
background \citep[see also][]{min02b,keshetetal02}.
According to results of \citet{min02b},
the \gr flux from the leptonic component scales with the 
group/cluster temperature less strongly than its hadronic 
counterpart \citep{mrkj01}. This means that it should dominate 
the \gr emission of small structures. In addition, as we shall
see in the ensuing sections, for conventional acceleration parameter 
values \gr  from \ic emission and \pnd can 
be of the same order even for a Coma-like cluster. 
Thus, the aimed detection of \gr emission from 
clusters may not necessarily reflect the CR hadronic 
component and this should be borne in mind
for a correct interpretation of the observational results.

With the objective of a correct diagnostic of future \gr observational
results, in this paper we compute the radiation spectrum and spatial
distribution of non-thermal emission from groups/clusters 
of galaxies. We inspect the case of
CR ions and electrons accelerated at structure shocks
as well as secondary electron-positrons ($e^\pm$).
However, we neglect the CRs that might originate at 
galactic wind termination shocks \citep{voahbr96} 
and radio galaxies.
The computed spectrum extends from hard X-ray (HXR) at 10 keV 
up to extreme \gr at 10 TeV. 
We examine the following emission processes: \ic and 
non-thermal bremsstrahlung (NTB) from shock accelerated
electrons, \ic emission from \epm and $\gamma$-rays from 
$\pi^0$-decay produced by CR ions.
We find that NTB emission is negligible throughout the spectrum.
\ic emission from shock accelerated electrons should dominate 
at HXR energies and be comparable to the flux from $\pi^0$-decay 
at \gr energies. 
Finally, \ic emission from \epm is typically below
these two components at all photon energies.
However, we show that all of these three emission components 
(\ie $e^-$, \epm and $\pi^0$) can potentially be 
individually measured by future \gr detectors with imaging 
capability by taking into account their peculiar 
spatial distribution and spectral properties. 
In particular we find that
\ic emission from shock accelerated CR electrons
dominates the emission at the outer regions of galaxy clusters,
provided that at least a fraction of a percent of the shock 
ram pressure is converted into CR electrons. Therefore, 
a clear detection of this component and of its spatial 
distribution will allow us direct probing of cosmic 
accretion shocks.

The paper is structured according to the following rationale:
The numerical models for both the large scale structure 
and the CR evolution are described 
in \S \ref{numtec.se}. The results on the non-thermal emission
are presented in \S \ref{res.se}, discussed in \S \ref{disc.se}
and finally summarized in \S \ref{summ.se}.

\section{The Numerical Model} \label{numtec.se}

The objective of this study is to investigate the spatial and
spectral properties of nonthermal \gr emission from 
CR electrons and ions accelerated at large scale structure 
shocks surrounding clusters/groups of galaxies. 
The task at hand requires accurate modeling of: 
(1) the large scale structure shocks where CR acceleration 
occurs; (2) the distribution within galaxy clusters of the 
baryonic gas which provides the CR targets for production 
of $\pi^0$ and secondary $e^\pm$; 
and (3) the spatial propagation and energy losses of 
the CRs. 
Furthermore, for a consistent description of the CR ions
it is important to account for (the acceleration due to) all the 
shocks that have processed the gas that ends up within the GCs 
themselves. That is because CR ions with momenta above 1 GeV/c and 
up to $\sim 10^{18}$ eV/c have a lifetime against energy losses longer 
than a Hubble time. 

This suggests that the formation of clusters and groups of galaxies 
be followed ab initio. Furthermore in order to reproduce meaningfully
the shocks where the CR acceleration occurs, this should be done 
within the ``natural'' framework of cosmological structure formation. 
For this reason we have carried out a numerical simulations of 
structure formation which follows the evolution of both the gas 
and dark matter in the universe starting from fully cosmological 
initial conditions. Details on the cosmological simulation 
are given below in \S \ref{cosmosim.se}.
In addition, with the numerical techniques described in the following 
sections (\S \ref{numcr.se}) we were also able to include 
the evolution of CR ions, electrons and 
secondary \epm by accounting explicitly for 
the effects of CR injection 
(at shocks and through p-p inelastic collisions), diffusive shock 
acceleration, energy losses and spatial propagation. 
Thus in addition to the gas distribution 
the simulation will also provide us with the information about
both the spatial and {\it momentum} distribution for each
CR components. In particular this information will be used in 
the \S \ref{res.se} to compute the spatial and
spectral distribution of nonthermal \gr emission
within gravitationally collapsed structures. Some of these results 
were already preliminarily presented in \citet{min02c,min02d}. 
Also, the simulation upon which the study is based, has 
been used by the author to study the contribution of \gr 
emission from CRs in the diffuse intergalactic medium 
to the cosmic \gr background \citep{min02b}.

In the following subsections we describe in further detail the 
salient features of the employed numerical techniques. These are
basically the same as those used in previous related studies 
\citep{mrkj01,mjkr01,min02b,min02d}, and the reader
familiar with them can skip directly to the 
result section in \S \ref{res.se}.

\subsection{Large Scale Structure} \label{cosmosim.se}

The formation and evolution of the 
large scale structure is computed by means of an
Eulerian, grid based Total-Variation-Diminishing
hydro+N-body code \citep{rokc93}. 
We adopt a canonical, flat $\Lambda$CDM cosmological model
with a total mass density $\Omega_m=0.3$ and a
vacuum energy density $\Omega_\Lambda= 1- \Omega_m= 0.7$.
We assume a normalized Hubble constant
$h\equiv H_0/100$ km s$^{-1}$ Mpc$^{-1}$ = 0.67 
\citep{freedman00} and a baryonic mass density, $\Omega_b=0.04$. 
The simulation is started at redshift $z\simeq 60$. 
The initial density field is homogeneous except for 
perturbations generated as a Gaussian random field 
and characterized by a power spectrum with a spectral index 
$n_s=1$ and ``cluster-normalization'' $\sigma_8=0.9$. 
The initial velocity field is then computed accordingly 
through the Zel'dovich approximation.

The choice of the computational box size
is a compromise between the need of a cosmological 
volume large enough to contain a satisfactory sample of
collapsed objects and numerical resolution requirements.
Ideally one would like to be able to produce rich clusters
of galaxies of the size of Coma cluster because they can be 
compared more easily with observations. 
However, in order for such massive objects to develop out of  
the initial conditions described above, the size of the 
computational box should be at least a few hundred \hinv 
Mpc in size.  
Although this could be achieved with Adaptive Mesh 
Refinement techniques, with the available computational 
resources and the code employed here the above box size would 
impose an unacceptably coarse spatial resolution.
Therefore the size of the computational box is set to 
$L=50$ \hinv Mpc. Given the relatively small box size the 
dimensions of the largest collapsed objects, with a
core temperature $T_x \simeq 2-4$ keV, are modest. 
Therefore, when in \S \ref{synmaps.se} and \S \ref{synspe.se} 
we make predictions for a Coma-size cluster 
(with $T_x \simeq 8$ keV)
the computed simulation results will be rescaled appropriately. 
The details of the rescaling procedure will be described in 
those sections.

Finally, the dark matter component is described by 256$^3$ particles
whereas the gas component is evolved on a comoving grid of 512$^3$ 
zones. Thus each numerical cell measures about 100 \hinv kpc
(comoving) and each dark 
matter particle corresponds to $2\times 10^9$ \hinv M$_\odot$.

\subsection{Cosmic-rays} \label{numcr.se}

In addition to the baryonic gas and dark matter, the simulation also
follows the evolution of three CR components, namely CR ions and
electrons injected at shocks and secondary \epm generated in p-p 
inelastic collisions of the CR ions with the gas nuclei.
This is achieved through the code COSMOCR fully described 
in \citet{min01,min02b}.

The dynamics of each of these CR component consist of the following
processes: injection, diffusive shock acceleration, energy
losses/gains and spatial propagation. In this section we describe 
how each of these processes is modeled numerically and how it is 
implemented in the simulation. 
We notice from the outset
that CRs are treated as passive quantities, meaning that their 
dynamical role is completely neglected both on the shock 
structure ({\it test particle limit}) and the gas dynamics. 

\subsubsection{Injection and acceleration at shocks} \label{injac.se}

The first step in order to compute CR injection and
acceleration at shocks, is to detect the shocks 
themselves. Here, as in previous studies, 
shocks are identified as converging flows
(${\bf \nabla \cdot v < 0}$) experiencing a pressure
jump $\Delta P/P$ above a threshold corresponding to a 
Mach number M=1.5. For the identified shocks both mass 
flux and Mach number are computed. 
This is simply done by evaluating the jump conditions
experienced by the flow as reproduced in the numerical 
simulation \citep[\eg][]{lali6}. 
Shocks are found both around filaments 
and groups/clusters.
In the latter case they show a rather complex structure
\citep[see \S \ref{rhovel.se} and]
[for a detailed description of cosmic shocks]{minetal00}.
An example of gas density distribution, velocity field and 
shock structure within a collapsed object is provided in 
Fig. \ref{velf.fig}, which illustrates the dominance of asymmetry
in the accretion flows and the existence of strong accretion
shocks at the outskirts region followed by weaker ones in the
inner regions. A detailed description of Fig. \ref{velf.fig}
is delayed until \S \ref{rhovel.se} and we shall now focus
on the description of the adopted CR injection/acceleration scheme.

As we shall see in the following, the mass flux through the shock 
and the shock Mach number determine two important 
quantities: respectively the CR injection rate and the 
shape of the accelerated distribution function.
Injection and diffusive
acceleration at shocks take place on spatial scales of 
the order of the particles diffusion length which is
\begin{equation}
\lambda_d(p) = 
1.1\; \left(\frac{E}{\mbox{GeV}}\right)
\left(\frac{B}{0.1\mu\mbox{G}}\right)^{-1}\;
\left(\frac{u_s}{10^2\mbox{Km\,s}^{-1}}\right)^{-1} 10^{-2} ~~~\mbox{pc}.
\end{equation}
where $E$ is the particle energy,
$B$ is the magnetic field strength, $u_s$ the shock speed 
and we have assumed for simplicity Bohm diffusion. As discussed in 
\citet[][see also \citealt{jre99}]{min01} to properly follow 
injection and acceleration at shocks one should {\it at least}
resolve not only scales $\sim \lambda_d(p)$ but also the shock 
structure including the subshock. 
Because such small scales can not be reproduced in the current 
simulation, injection and diffusive acceleration are 
{\it not} directly simulated but simply modeled.
In particular, given the mass flux across a shock, 
the assumed injection model determines the fraction of particles
that are converted into CRs. Then, the acceleration model determines
how the injected CRs should end up distributed in momentum space.

%
%
For the injection of CR ions we follow the thermal 
leakage prescription \citep{elei84,quest88,kajo95}. That is,
the post-shock gas is assumed to thermalize to a quasi-Maxwellian
distribution function characterized by a temperature $T_{\rm shock}$. 
The ions/protons in the high energy tail of 
such distribution are assumed to be 
able to leak back upstream and undergo the
acceleration mechanism. The momentum threshold for such injection 
is set to $p_{inj}= 2 c_1 m_p \sqrt{k_BT_{\rm shock}/m_p}$,
where $m_p$ is the proton mass and $c_1$ is a control parameter
\citep{kajo95}. We adopt $c_1\geq 2.5$ which implies that 
a fraction $\sim 10^{-4}$ of the particles crossing the shock
be injected in the acceleration mechanism. In terms of 
shock ram pressure, the 
pressure borne by the CR ions is a small fraction for weak shocks
($M\leq 3$) and reaches 30 \% for moderately strong 
shocks ($M\sim 4-6$). Given our simplified injection prescription 
for very strong shocks ($M\geq 10$) this fraction can approach 
unity or so. Therefore, for consistency in these cases the fraction
of injected particles is renormalized so that the CR pressure 
is always limited below 40\% of the shock ram pressure.

As for electrons (primary $e^-$), we simply assume that the 
ratio between injected CR electrons and ions at
relativistic energies be given by a parameter $R_{e/p}$. 
The introduction of this parameter
simplifies the treatment of this process, which is quite complex 
and yet to be fully understood
\cite[see, \eg][for a discussion on this issue]{elbeba00}.
Observationally, for Galactic CRs this ratio appears to be in the 
range $1\times 10^{-2}-5 \times 10^{-2}$ \citep{muta87,mulletal95,alpego01}.
Also, based on EGRET observational upper limits on the \gr flux from
nearby clusters \citep{sreeku96,reimer99,reposrma03}, 
for a CR ion injection efficiency as assumed here
\cite{min02b} constrained $R_{e/p}\leq 2 \times
10^{-2}$. Therefore, in the following we set $R_{e/p}=10^{-2}$
and all of our results concerning primary $e^-$ are based on this 
assumption.

%
%
CR ions and electrons injected as described above, are quickly
accelerated to energies much higher then the thermal average
through the diffusive shock acceleration mechanism. As already
pointed out, given the prohibitively short length scales over 
which the acceleration process takes place, its direct inclusion 
in the simulation is not possible. In addition, for a typical 
shock with compression ratio close to four the acceleration time
scale
\begin{equation} 
\tau_{acc}(p)  =  
 21.1\; \left(\frac{E}{\mbox{GeV}}\right)
\left(\frac{B}{0.1\mu\mbox{G}}\right)^{-1}\;
\left(\frac{u_s}{10^2\mbox{Km\,s}^{-1}}\right)^{-1}\; \mbox{yr}
\end{equation}
is much shorter than the simulation dynamical time 
(computational time-step). 
Thus, as in previous studies, we assume that the injected CRs 
are accelerated {\it instantly} and distribute them in momentum
according to a power-law distribution extending from injection 
up maximum energy, \ie
\begin{equation}  \label{finj.eq}
f(p)=f_0 \left(\frac{p}{p_{inj}}\right) ^{-q},
~~~~~~~~ p_{inj}\leq p \leq p_{max}.
\end{equation}
For CR ions $f_0$ is determined through the aforementioned 
thermal leackage injection model and for 
CR electrons it is such that at relativistic ($\geq$ GeV) energies their
ratio to accelerated protons is $\simeq R_{e/p}$.
According to the test particle limit adopted here,
the log-slope of the distribution function is related to the 
shock Mach number, $M$, as 
$q=3(\gamma +1)/[2(1 - M^{-2})]=4/(1-M^{-2})$ for $\gamma=5/3$. 
This implies relatively flatter distribution functions for 
stronger shocks.

Based on cluster/group properties, 
we found it appropriate to follow CR ions between momenta 
$p_{min}= 10^{-1}$GeV/c and $p_{max}= 10^6$GeV/c. In fact,
CR ions below $p_{min}$ quickly loose energy due to Coulomb 
losses and beyond $10^{7}$ GeV/c their confinement within 
cosmic structures becomes difficult \citep{voahbr96,cobl98}.
CR electrons are followed between $p_{min}=15$ MeV and 
$2\times 10$ TeV. The latter is a reasonable value for 
the maximum energy to which CR electrons are shock accelerated, 
provided a magnetic field of order $0.1 \mu$G.

\subsubsection{Secondary Electrons and Positrons}

In addition to ions and electrons injected and accelerated
at shocks, CR electrons and positrons are also produced 
in hadronic collisions of CR ions with the nuclei of the 
intra-cluster gas. The parent CR ions are those computed in 
the simulation. Secondary $e^\pm$ are generated 
in the decay of charged muons according to
\begin{eqnarray} 
\mu^\pm  \rightarrow e^\pm + \nu_e (\bar{\nu}_e) + \bar{\nu}_{\mu}(\nu_{\mu})
\end{eqnarray}
\noindent
which in turn are produced in the following reactions
\begin{eqnarray} 
p + p & \rightarrow \pi^\pm +  X,~~ & \pi^\pm  \rightarrow \mu^\pm + \nu (\bar{\nu}) \\
p + p & \rightarrow K^\pm + X, ~~ & K^\pm  \rightarrow \mu^\pm +\nu (\bar{\nu})\\
& & K^\pm  \rightarrow \pi^0 + \pi^\pm
\end{eqnarray}
\noindent
Additional \epm are produced in 
cascades analogous to these by the following 
interactions: $p+$He, $\alpha +$H and $\alpha +$He. 
These are included by assuming a helium number fraction
of 7.3\% for the background gas and a ratio
$(H/He) \simeq 15$ at fixed energy-per-nucleon 
for the CRs \citep{medrel97}. 
The distributions of the \epm thus generated are obviously related
to those of the parent CR ions. In particular, if the latter are
distributed according to a power law, so are also the former.

\subsubsection{Energy losses and spatial propagation}

The numerical treatment of the CR dynamics is completed by accounting 
for both spatial transport and energy losses/gains as described by 
the diffusion-convection equation \cite[\eg][]{skill75a,skill75b}.
In order to compute a numerical solution to this equation, 
we define a grid in 
momentum space by dividing it in $N_p$ logarithmically 
equidistant intervals ({\it momentum bins}). 
For each mesh point of the spatial grid, ${\bf x_j}$, 
the CR distribution function is then described by the 
following piecewise power-law \citep{jre99,min01}
\begin{equation} \label{distf.eq}
f({\bf x}_j,p) = f_j({\bf x}_j) \, p^{-q_i({\bf x}_j)},
~~~~~~~~ 1<p_{i-1} \le p \le p_i,
\end{equation} 
where $p_i ... $ are the bins' extrema. 
Spatial propagation and energy losses of the accelerated CRs are 
then followed by solving numerically a diffusion-convection
equation that has been multiplied by `$4\pi\,p^2$' 
and integrated over each momentum bin. 
Written in the comoving coordinates system adopted for the 
other simulated hydrodynamic quantities, 
such equation reads \citep{min01}
\begin{equation} \label{numden.eq}
\frac{\partial n({\bf x_j},p_i)}{\partial t} =
- \frac{1}{a}\; {\bf \nabla\cdot } \left[ {\bf u}\,n({\bf
x_j},p_i)\right] + 
 \left\{ \left[ \left(\frac{\dot{a}}{a}+ 
\frac{1}{3a}\, {\bf \nabla \cdot u}\right)\: p\,+ b_\ell(p) \right] 
\;4\pi p^2 f({\bf x_j},p_i)\right\}_{p_{i-1}}^{p_i} 
+Q({\bf x_j},p_i). 
\end{equation}
Here $n({\bf x},p_i) = 
\int_{p_{i-1}}^{p_i} 4\pi\,p^2 \;f({\bf x_j},p)\; dp$ is 
the comoving number density of CR in the i-th momentum bin
and $Q({\bf x_j},p_i)$ is a comoving source term, $i({\bf x_j},p)$, 
describing either shock injection or \epm generated in 
p-p collision and integrated also over the i-th momentum bin.
In addition, $a$ is the cosmological expansion factor, $\dot{a}/a=
H(z)$ is the Hubble parameter defining the cosmic expansion rate. 
Thus the first part of the second term in eq. (\ref{numden.eq}) 
describes adiabatic losses/gains. 
Finally, $b_\ell(p)$ represents energy losses due to
Coulomb collision, bremsstrahlung, \ic and synchrotron 
emission for electrons and $e^\pm$; and Coulomb and p-p inelastic
collision for ions. All of them are fully described in \citet{min01}.
In eq. (\ref{numden.eq}) ${\bf u}$ is the velocity field of the 
baryonic gas to which the CRs are assumed to be coupled through 
a background magnetic field. Thus once accelerated the CRs are 
passively advected with the gas flow. 
Notice that eq. (\ref{numden.eq}) does not include the diffusion 
term present in the ordinary diffusion-convection equation.
This is because, as pointed out in 
\citet[][see also \citealt{jre99}]{min01} for typical flows
in the simulation with $u \sim$ a few 100 km $s^{-1}$
diffusion over spatial scales $\Delta x \simeq $ 100 kpc is much 
slower than advection and can be neglected away from shocks.
This is true for values of the diffusion coefficients typically 
assumed in the literature \cite[see, e.g.,][]{voahbr96,bbp97}. 
The value
inferred for these diffusion coefficients is based on the assumption
that the magnetic field fluctuations are described by a turbulent
spectrum and that the total magnetic field energy density is that 
characteristic of $\mu$G strong magnetic fields.

Integrating eq. (\ref{numden.eq}) is sufficient for an accurate 
treatment of the ionic component. In this case, after advancing the
solution to eq. (\ref{numden.eq}) one time-step, both 
$f_j({\bf x}_j)$ and $q_i({\bf x}_j)$ defining the CR 
distribution function in eq. (\ref{distf.eq}) are computed based 
on the updated values of $n({\bf x_j},p_i)$ and by assuming that
$f({\bf x}_j,p)$ is continuous at each bin interface.

For CR electrons, however, 
severe energy losses and distribution cutoffs render 
their numerical treatment much more delicate. For this 
reason for this component in addition to $n({\bf x},p_i)$ 
we also follow the correspondent bin kinetic energy 
$\varepsilon({\bf x},p_i) = 
4\pi \int_{p_{i-1}}^{p_i}\,p^2 \;f({\bf x},p_i)\,T(p)\; dp $. 
Here $T(p)=(\gamma -1)\,m_e\,c^2$ is the particle kinetic energy and
$\gamma = [1-(v/c)^2]^{-1/2}$ is the Lorentz factor.
The equation describing the evolution of 
$\varepsilon({\bf x_j},p_i)$ is obtained in analogy to eq. 
(\ref{numden.eq}), by integrating over the i-th 
momentum bin the same diffusion-convection equation (more properly
a kinetic equation) that has been multiplied by $T(p)$.
In comoving units this reads 
\citep{min01}
\begin{equation} \label{enden.eq}
\frac{\partial \varepsilon ({\bf x_j},p_i)}{\partial t} =
-\frac{1}{a}\;  
{\bf \nabla\cdot } \left[ {\bf u}\,\varepsilon({\bf x_j},p_i)\right] + 
\left[ b(p) \;4\pi \;p^2\; f({\bf x_j},p)\,T(p) \right] _{p_{i-1}}^{p_i} -
\int_{p_{i-1}}^{p_i} b(p)
\;\frac{4\pi c p^3 f({\bf x_j},p)}{\sqrt{m_ec^2+p^2}}\;dp
+ S({\bf x_j},p_i),
\end{equation}
where $S({\bf x_j},p_i) = 4
\pi\;\int_{p_{i-1}}^{p_i}i({\bf x_j},p_i)\,p^2 T(p)\, dp$,
$b(p)$ now includes both adiabatic loss terms and those described 
by $b_\ell(p)$, and the third term on the right hand side
includes contributions from CR pressure work 
and sink terms. The evolution of the CR electrons is 
obtained by integrating numerically eq. (\ref{numden.eq}) 
and (\ref{enden.eq}) through a semi-implicit 
scheme described in \cite{min01,min02b}. The slope 
of the distribution function at each grid point and 
momentum bin is then determined self consistently by the values of 
$n({\bf x_j},p_i)$ and $\varepsilon({\bf x_j},p_i)$.

However, for the simulated CR electrons and \epm with momenta 
between $10^2$ GeV/c and 20 TeV cooling time due to \ic losses,
$\tau_{cool}\sim 10^5 - 10^7$yr, is much shorter than the typical 
computational time-step, $\Delta t\sim 10^7-10^8$yr.
Thus, because ${\bf u}\tau_{cool} \ll 
{\bf u} \Delta t \leq \Delta x$ (Courant condition),
these particles only exist in grid cells where 
the source term $i({\bf x_j}, p) \neq 0$ and will not 
propagate outside it.
Thus, as discussed in \citep{min02b}, in this case it is 
computationally advantageous and numerically correct to 
take directly the steady state solution to eq. (\ref{numden.eq}) as
\begin{equation}
n({\bf x_j},p_i) = 4\pi\int_{p_{i-1}}^{p_i}\,p^2 \; 
f_{c}({\bf x_j},p_i)\; dp  =
- 4\pi \int_{p_{i-1}}^{p_i} 
\frac{dp}{b(p)}\; \int_{p}^\infty \; \rho^2 \,i({\bf x_j},\rho)\, d\rho
  \label{sss.eq}
\end{equation}
\noindent
[a similar expression holds for eq. (\ref{enden.eq})].
Notice that $n({\bf x_j},p_i)$ refers to the cell-volume averaged
number density of CRs.
Physically, the expression in eq. (\ref{sss.eq}) represents
a summation of all the individual populations of CR electrons within 
the grid zone ${\bf x_j}$ that, as they emerge from the acceleration 
region, and are being advected away from the shock, develop a 
cut-off due to energy losses.
For what follows in the next sections it is relevant to 
notice that for an injection spectrum $i(p) \propto p^{-s}$
and for the case of energy losses dominated by \ic emission, 
$b(p)\propto p^2$, the steady state distribution given in eq. 
(\ref{sss.eq}) implies a cell-volume averaged distribution function
$\langle f(p) \rangle \propto n({\bf x_j},p_i)/\Delta x^3 \propto p^{-(s+1)}$.

Finally the CR ion distribution function is mapped by 4 momentum 
bins. For the electrons and \epm we used 5+1 momentum bins. 
The first five are logarithmically equidistant and cover a 
momentum range between $p_{min}=15$ MeV up to $p_{2}= 10^2$ GeV. 
The last electron momentum bin stores the CR electron
steady-state distribution function between $p_{2}= 10^2$ GeV
and 20 TeV given by eq. (\ref{sss.eq}).
Our tests indicate that increasing the number of momentum 
bins for either CR component (ions and electrons) does not 
affect the results.

\section{Results} \label{res.se}

In the following sections we present results based on the simulation
just described on the emission of high energy radiation from 
CRs in groups/clusters of galaxies.
In particular, we inspect the spectra, radial 
dependence and synthetic images for various emission 
mechanisms and for each computed CR population.
For the purpose we excise out of the simulation box regions
containing individual virialized objects. For each of them we
compute the nonthermal radiation produced through emission processes
described below, by using the simulated hydro (density) and 
CR (distribution functions) quantities. For simplicity in the 
following we focus only on a couple of simulated virialized 
objects, although the studied radiation properties
are independent of our peculiar choice. Finally, in \S \ref{radspe.se}
and \S \ref{lumvol.se} we describe the qualitative properties 
of the nonthermal radiation whereas in \S \ref{synmaps.se} and 
\S \ref{synspe.se} we will attempt quantitative predictions for a 
Coma-like cluster of galaxies. 

\subsection{Density and Velocity Structure} \label{rhovel.se}
\begin{figure}
\caption{\textit{Color image showing the density distribution
on a plane passing through the center of the collapsed object. 
It is in units of cm$^{-3}$ and the various level correspond to
values indicated by the colorbar on the right of each panel. 
Arrows describe the velocity field on the same plane. 
Their number has been reduced by a factor nine for clarity purposes.
Finally, blue contours indicate the isolevel of compression 
(${\bf \nabla \cdot v } < $0) where shoks occur. Narrow countour 
features correspond to location of strong shocks.
Left and right panel correspond to X-Y and Z-X coordinate planes
respectively. Axis scales are measured in \hinv Mpc.
}}
\label{velf.fig}
\end{figure}

First we consider the largest
virialized object in the computational box.
This is characterized by an X-ray core temperature 
$T_x\simeq 4$ keV. 
The object is partially described in fig. \ref{velf.fig} 
where the two panels show two dimensional cuts across the 
coordinate planes X-Y and Z-X respectively of 
the density distribution (in color scale),
the velocity field (arrows) and the 
location of shocks (contours). 
The gas density is $n_c \sim 6 \times 10^{-3}$
cm$^{-3}$ at the core and drops by about 2 orders of magnitudes
at a distance of a few Mpc. 
The tree-dimensional velocity characterizing the accretion 
flows is typically of order of 10$^3$ km s$^{-1}$. 
The velocity field converges toward the mass concentration
and becomes quite complex close to it.
As described in more detail in \citet{minetal00}, this is due to 
the large asymmetry of accretion flows. Noticeably, stream along
filaments, which carry more momentum, reach closer to the 
central regions before being shocked. 
As a result the ensuing shock structure is also complex.
The largest shocks are found up to $\sim 5$ \hinv Mpc from
the core of the collapsed object. Additional shocks are 
present at intermediate distances, especially along filaments
as already pointed out. These shocks have Mach numbers 
that range from up to $\sim$ 100 in the most outlining regions,
to $3-10$ along filaments. Close to the high temperature
core some discontinuities are also found. However, as it is 
apparent from the broad features of the isocontours, these
are weak, tran-sonic shocks. 

For the selected object 
the energy in CR ions is $E_{CR} \simeq 6 \times 10^{61}$
erg in the inner 1.5 Mpc and corresponds to about 
26\% of the thermal pressure within the same volume.  
We notice the presence in the neighborhood of the collapsed 
objects of a small structure which is 
is marginally visible above and to the right of it, 
in the left and right panels of 
fig. \ref{velf.fig} respectively. 
As we shall see in the following, 
this will affect (particularly, will increase) the computed 
radiation spectrum at large radii. 

\subsection{Radiation Spectrum} \label{radspe.se}

\begin{figure}
\includegraphics[width=0.50\textwidth ]{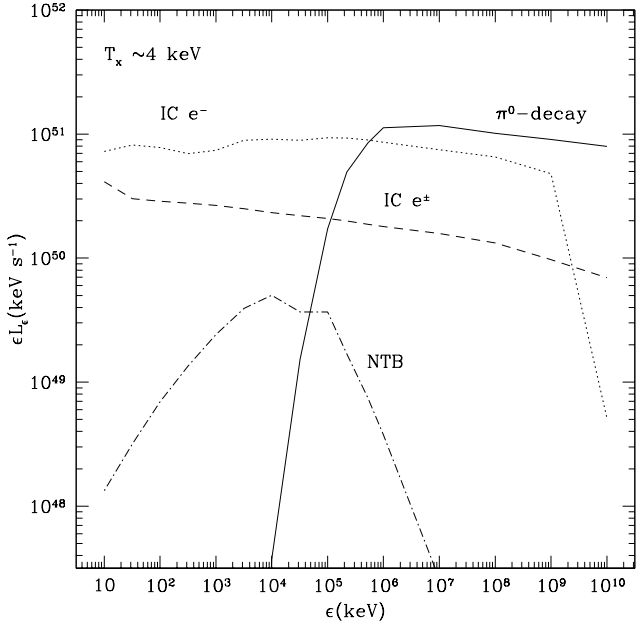}
\caption{\textit{Radiation spectrum extending 
from 10 keV up to 10 TeV produced by the 
following emission processes: \ic (dotted) and 
NTB (dash-dotted) from shock accelerated
CR electrons, \ic emission from \epm (dashed)
and $\gamma$-rays from $\pi^0$-decay produced by 
CR ions (solid).}}
\label{spectr.fig}
\end{figure}

In fig. \ref{spectr.fig} the quantity 
`$\varepsilon \; L_\varepsilon$' in units `keV s$^{-1}$' associated
with the selected object is plotted as a function of photon 
energy, $\varepsilon$.
`$L_\varepsilon $' is the volume integrated spectral power and is defined as 
\begin{equation} \label{phlum.eq}
L_\varepsilon (\varepsilon,R) = 
\int_{\rm V(R)} dV \; j_\varepsilon(\varepsilon) 
\end{equation}
where $j_\varepsilon(\varepsilon)$ is the spectral emissivity in units
`keV cm$^{-3}$ s$^{-1}$ keV$^{-1}$' and $V(R)=4\pi R^3/3$ is an 
integration volume defined by a radius $R$. Here we set $R= 5$ 
\hinv Mpc in order to account for
the shock accelerated electrons which are typically located 
away from the cluster center (see fig \ref{velf.fig}).
We consider the following emission processes: 
$\pi^0$-decay (solid line), \ic emission from shock 
accelerated electrons (dotted line) and from secondary e$^\pm$ 
(dashed line) scattering off CMB photons, 
and NTB from shock accelerated electrons (dot-dashed line). 
Thermal X-ray photons, of energy $\varepsilon_X$,
can be neglected as seeds for IC 
emission of $\gamma$-rays with respect to CMB photons 
to first order approximation. In fact, although the number of 
available IC scattering electrons is larger by a factor 
($\langle \varepsilon_X \rangle / 
\langle \varepsilon_{CMB} \rangle)^{(q-3)/2} $ this is not sufficient
to compensate for the lower number of seed photons (especially for
flat CR distributions such as those produced at accretion shocks) and for 
the IC cross-section suppression due to Klein-Nishina effects which enter 
for $x\simeq 1.7 [ \varepsilon({\rm GeV}) \varepsilon_X({\rm keV}) ]^{1/2}
\geq 1$ \citep[see also, \eg][]{enbi98,bede02}.

According to fig. \ref{spectr.fig}, 
both emission from $\pi^0$-decay and \ic from primary \cre 
contribute significantly to the \gr emission above 100 MeV.
The actual proportion between these two components is 
determined by two factors: (a) The ratio of accelerated
electrons and ions at relativistic energies, which has never 
been measured for cluster shocks. It is 
represented by the parameter $R_{e/p}$ and was set to $10^{-2}$
in the current study. (b) The temperature of the group/cluster,
from which the \gr luminosity from $\pi^0$-decay and \ic emission
depend differently. In fact, for both processes
the temperature dependence is expected to be of power-law type, but 
with a moderately steeper index for the hadronic emission component 
\citep{mrkj01,min02b}. Thus the relative strength of these two
contributions as presented in fig. \ref{spectr.fig} is only 
indicative at this stage. However, it is within the objectives of 
the present paper to investigate possible strategies for their
experimental determination.
At HXR and soft \gr energies below 10 MeV, \ic from primary \cre
dominates the spectrum although the contribution from
\epm is not negligible.
However, as we shall see in the next sections, 
this source of emission should also be detectable 
at HXR energies owing to its different spatial distribution 
with respect to the primary \cre.

Noticeably the radiation spectra of these 
three components are rather ``flat''. This indicates
that the primary CR ions and $e^-$'s were 
generated in shocks of at least moderate strength, 
\ie $M\geq 4$. Indeed, for the hadronic component, 
the \gr emissivity 
from a CR ion distribution $f(p)\propto p^{-q_p}$, 
is $j_\varepsilon(\varepsilon) 
\propto \varepsilon^{-(4q_p-13)/3}$ \citep[\cf][]{masc94}.
Thus, the case of an emission spectrum such that
$\varepsilon L_\varepsilon \propto \varepsilon j_\varepsilon
\propto \varepsilon^0 $, implies a CR ion 
distribution with $q_p\simeq 4$.
As for both primary electrons and secondary \epm, 
because of the severe energy losses suffered
by these particles, the emission spectrum must be related to their
steady state distributions.
As pointed out at the end of \S \ref{numcr.se}, this is characterized
by a steady-state log-slope, $q_{ss}=q_{source}+1$, 
that is steeper by one unit with respect 
to the source spectrum. The produced \ic emissivity is of the form
$j_\varepsilon(\varepsilon)
\propto \varepsilon^{-(q_{ss}-3)/2}$ which satisfies the condition 
$\varepsilon j_\varepsilon \propto \varepsilon^0 $
for $q_{ss} \sim 5$ and, therefore, $q_{source} \sim 4$. 
For primary \cre this straightforwardly requires that these CRs
were accelerated at strong shocks. 
For secondary \epm with a source term of the form
$f_{e^\pm}(\varepsilon_{e^\pm}) \propto \varepsilon_{e^\pm}
^{-4(q_p-1)/3}$ \citep[\cf][]{masc94}, it requires that the
parent CR ions were accelerated at strong shocks and that 
their log-slope distribution be $q_p\simeq 4$.

Finally, NTB is negligible for all purposes. Notice that 
latter component includes also the contribution from 
trans-relativistic electrons. These manage to propagate away
from the acceleration region (\ie a shock) for a short time 
before being re-absorbed into the thermal
pool due to Coulomb losses. This trans-relativistic
component enhances the NTB emissivity with respect to 
that produced by the bulk of the relativistic electrons. 
In fact, it dominates the emission 
below 10 MeV causing a sort of bump
in the shape of the spectrum about this photon energy
(dot-dashed line in fig. \ref{spectr.fig}).

\subsection{Luminosity Volume Dependence} \label{lumvol.se}

We now inspect the spatial distribution 
of the emissivity associated with the various
processes presented in the previous section. 
As anticipated there, such information allows 
in principle a measurement of each individual 
emission component (except NTB).
\begin{figure}
\includegraphics[width=0.50\textwidth ]{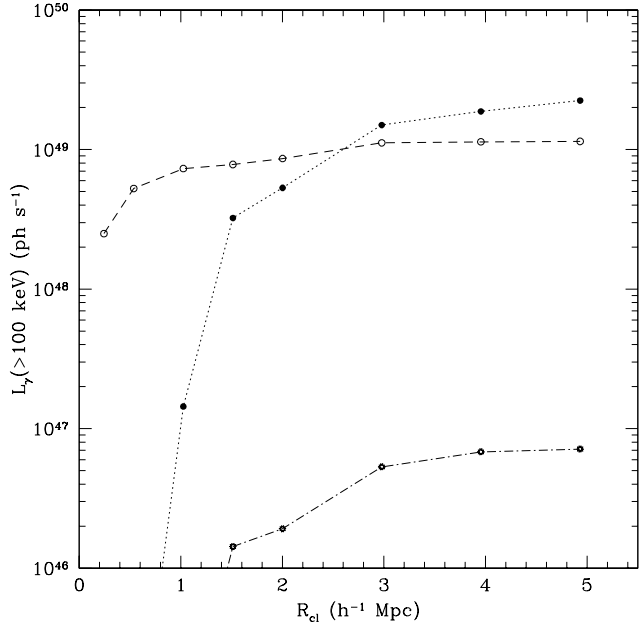}
\caption{\textit{Radial dependence of the integrated
photon luminosity above 100 keV for the
following emission processes: \ic (dotted) and 
NTB (dash-dotted) from from shock accelerated
CR electrons, \ic emission from \epm (dashed).}}
\label{rads.fig}
\end{figure}
With this in mind, in fig. \ref{rads.fig} we show the radial 
dependence of the integrated 
photon luminosity above 100 keV, that is
\begin{equation} \label{iphlum.eq}
L_\gamma({\rm >100~ keV, R}) =
\int_{\rm > 100~ keV} L_\varepsilon(\varepsilon,R)
 \; d\varepsilon
\end{equation}
where $R$, the radius defining the integration volume, 
is now a varying parameter.
Obviously, at these low energies there is 
no contribution from the CR ionic component. 
Thus the diagram contains only three curves illustrating \ic 
emission from primary \cre (dotted) and secondary \epm
(dashed) and, for completeness, NTB from \cre (dot-dashed).
This plot clearly shows how the radiation emitted by primary 
\cre and secondary \epm originates in spatially separate regions. 
In fact, the latter component saturates quickly 
within the central Mpc and dominates the total 
photon production there. On the contrary, the contribution 
from the former component becomes significant 
only at a distance of about 1.5 \hinv Mpc from the cluster 
center and keeps increasing up to several Mpc from there. 

\begin{figure}
\includegraphics[width=0.50\textwidth ]{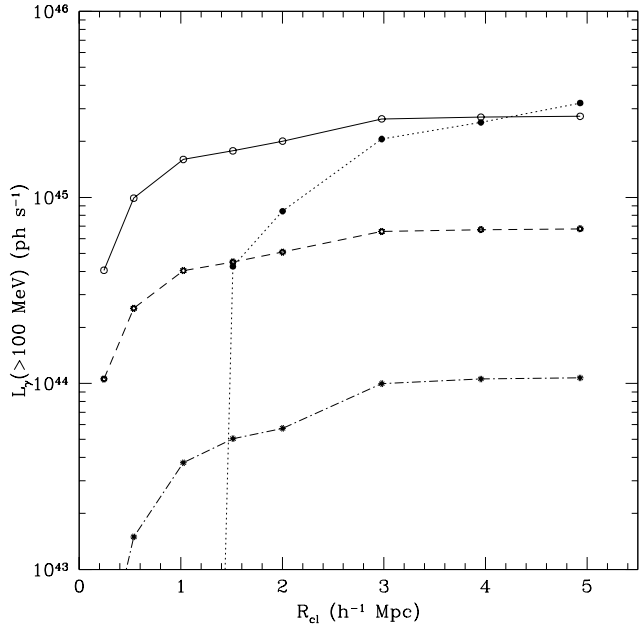}
\caption{\textit{Radial dependence of the integrated
photon luminosity above 100 MeV for the
following emission processes: \ic (dotted) and 
NTB (dash-dotted) from shock accelerated
CR electrons, \ic emission from \epm (dashed)
and $\pi^0$-decay (solid).}}
\label{radh.fig}
\end{figure}
The situation is analogous at \gr energies. This is illustrated 
in fig. \ref{radh.fig} where the integrated photon luminosity 
above 100 MeV, $L_\gamma ({\rm >100~ MeV, R})$ - defined by an 
equation analogous to
eq. (\ref{iphlum.eq}) - is plotted versus radial distance. 
Here it is the emission from $\pi^0$-decay (solid line) 
that saturates at relatively short distances from the cluster 
center and dominates the photon emission in the 
inner regions.  As before, \ic emission from primary 
\cre (dotted line) reaches a substantial level only outside a 
distance $\sim 1.5$ \hinv Mpc. 
\ic emission from \epm (dashed line) is now
unimportant and, as in the previous case, NTB 
from primary \cre (dot--dashed line) is completely negligible.

The radial dependence of the various radiation components presented 
in fig. \ref{rads.fig} and \ref{radh.fig} reflects both the spatial 
distribution of the emitting particles 
as well as the nature of the emission process. Thus, on the one hand, 
both \epm and $\pi^0$ are produced at the highest rate 
in the densest regions where both the parent CR ions and target
ICM nuclei are most numerous. Consequently, \epm \ic and \pnd 
emissivities are strongest in the cluster inner regions and 
quickly fade toward its outskirts. 
Notice that in the case of fig. \ref{rads.fig} and \ref{radh.fig}
this behavior is slightly altered by
the presence of a nearby object (similar to the case in
fig. \ref{maps.fig} below) 
which causes these integrated photon luminosities 
to slowly grow even after few core radii instead of leveling out.

On the other hand,  because of severe energy losses, 
the emitting high energy primary \cre are only found 
in the vicinity of strong shocks where they are being accelerated. 
Notice that in this case the shocks must be strong so 
that enough particles are accelerated to high energies 
to produce appreciable emission. Thus,
because the strongest shocks are located at the cluster
outskirts rather than at its core where the ICM temperature is 
already high \citep[fig. \ref{velf.fig}; also \cf][]{minetal00}, 
the spatial distribution of the 
emissivity is now reversed with respect to the previous case. 
Notice that, in line with this analysis, the drop of \ic  
emissivity (from primary \cre)
toward the inner regions is slightly more abrupt
at \gr energies (fig. \ref{radh.fig}) 
than at HXR energies (fig. \ref{rads.fig}). 

\subsection{Synthetic Maps} \label{synmaps.se}
\begin{figure}
\includegraphics[width=0.75\textwidth ]{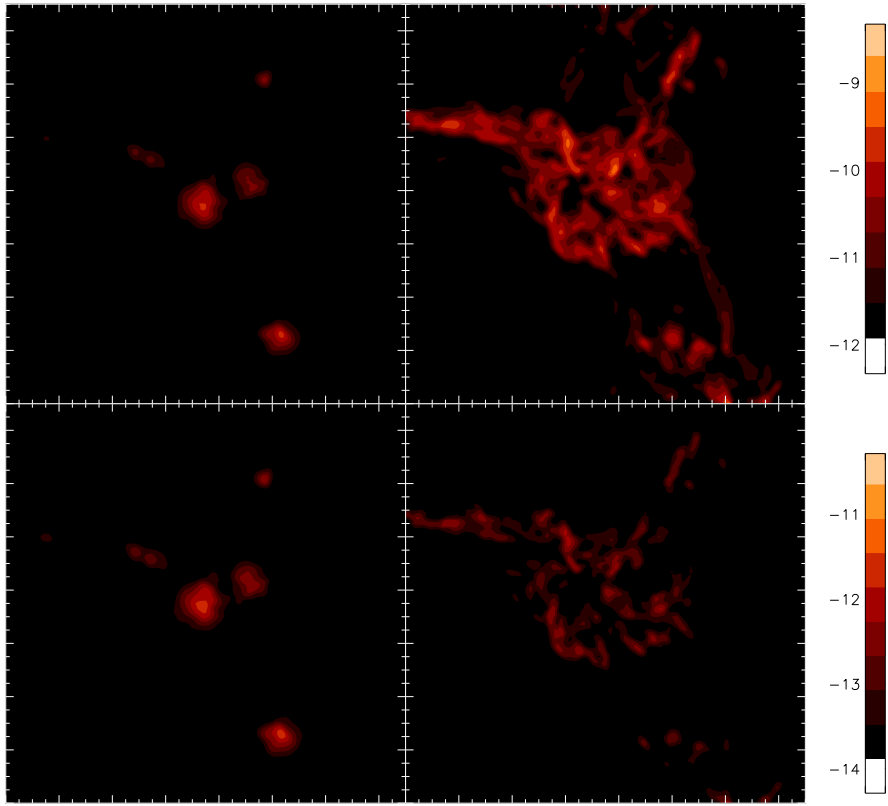}
\caption{\textit{Synthetic map 
of the integrated photon flux above 100 keV (top) and 
100 MeV (bottom) 
in units ``ph cm$^{-2}$ s$^{-1}$ arcmin$^{-2}$'' from
\ic emission by secondary \epm (top-left), 
primary \cre (top-right, bottom-right),
and \pnd (bottom-left). Each panel measures 15 \hinv Mpc on a side.}}
\label{maps.fig}
\end{figure}
\begin{figure}
\includegraphics[width=1.\textwidth ]{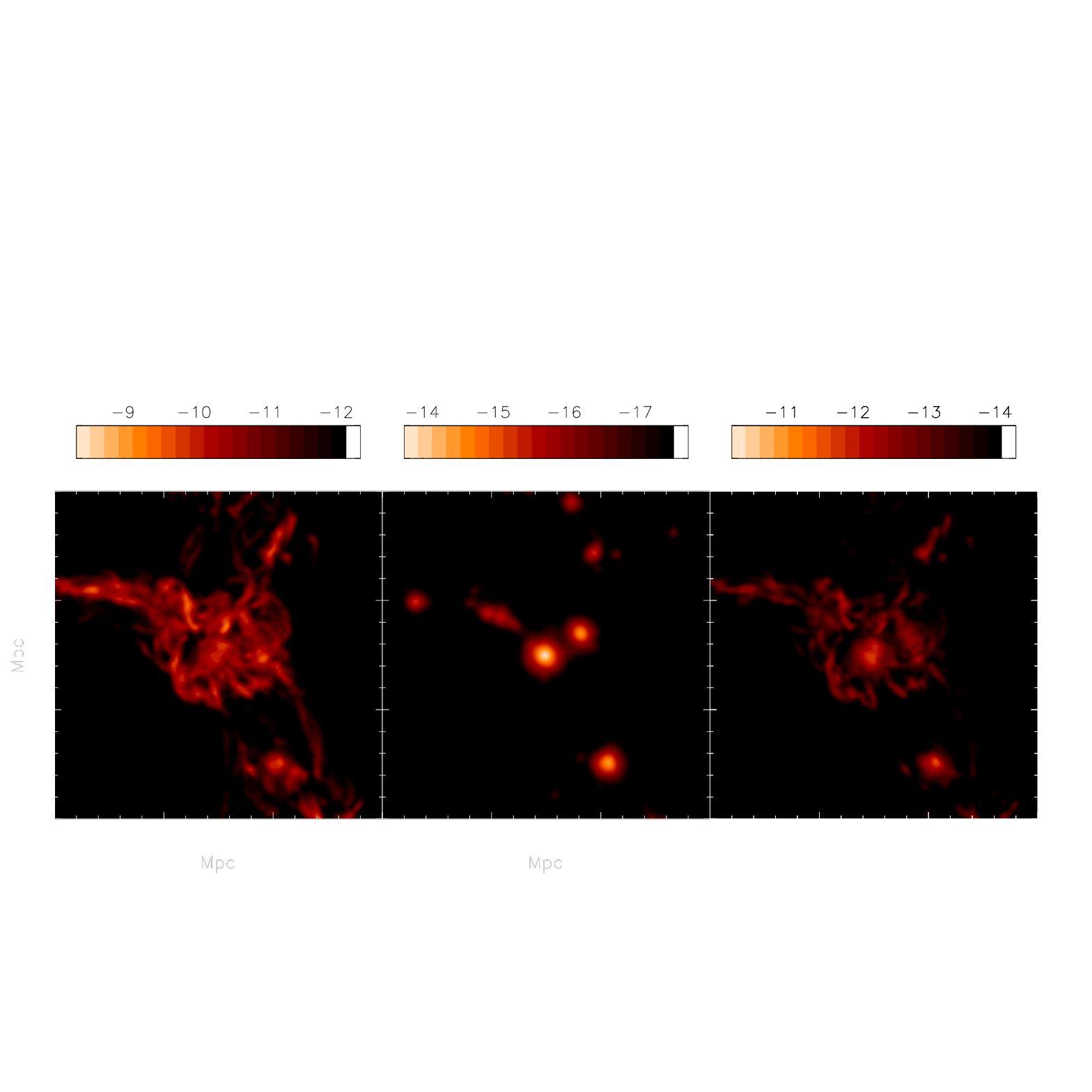}
\caption{\textit{Left, Right: Synthetic map of the total photon
flux above 100 keV (left) and 100 MeV (right) 
in units `ph cm$^{-2}$ s$^{-1}$ arcmin$^{-2}$'. Left panel includes
\ic emission from primary \cre, secondary \epm and NTB.
Right panel includes emission due to \pnd, \ic from primary \cre, 
secondary \epm and NTB (c).
Center: Synthetic map of the bolometric X-ray emission 
from thermal bremsstrahlung in units `erg cm$^{-2}$ s$^{-1}$ 
arcmin$^{-2}$'. Each panel measures 15 \hinv Mpc on a side.}}
\label{maph.fig}
\end{figure}
In this section we present synthetic maps produced by non-thermal 
emission processes. These maps were obtained upon simple integration 
along the line of sight of the photon emissivity in the thin 
plasma approximation. We use them to study the morphology of the 
emitting region. In addition we investigate which emission component
contributes most to the radiation spectrum when the latter is extracted 
from different spatial regions. 
Unlike plots in fig. \ref{rads.fig} and \ref{radh.fig}, 
using synthetic maps will automatically account for projection effects. 

We will attempt to relate 
the present analysis to the case of Coma cluster which,  
for its large size and relative vicinity, is likely the best source
candidate for the detection of non-thermal, high energy radiation.
Our calculations are now based on a different collapsed object 
extracted as before from the simulation box but relatively more
isolated (and therefore better suited for the current purpose)
than the previous one.
It has a temperature of a few keV, a core density of 
$2\times 10^{-4}$ cm$^{-3}$. 
This object is smaller than Coma cluster and for that reason 
we will renormalize the emission components as follows.\footnote{
Notice that this ``renormalization'' procedure can be applied to 
any other cluster for which sufficient observational data are
available.} 
As for the hydro quantities we retain the temperature and density 
profiles produced in the simulation but renormalize them by 
(multiplying each of them by) the ratio of the observed and 
simulated core values. The following observed values for Coma 
cluster are taken: a temperature
$T_x=8.2$ keV \citep{arnaudetal01} and a gas density of 
$n_{gas} \simeq 3\times 10^{-3}$ cm$^{-3}$. The thermal energy 
within 1 Mpc computed according to temperature and density
thus renormalized corresponds to $1.6\times 10^{63}$ erg.  

For the \ic emission from primary \cre we have used the recent finding 
of \cite{min02b} indicating that the \gr flux due to this process
scales with the X-ray temperature as $F_\gamma \propto T_x^{2.6}$. 
The total number of secondary $e^\pm$, $N_{e^\pm}$, and the associated 
\ic flux $F_{IC}^{\pm}$, are normalized by assuming that these 
particles are responsible
for producing Coma radio halo through emission of synchrotron radiation. 
For a given measured radio flux, $S_{sy}$, and an assumed average 
ICM magnetic field strength, $\langle B \rangle$, 
the total number of emitting particles scales as 
\begin{equation} \label{norm1.eq}
F_{IC}^{\pm} \propto N_{e^\pm}
\propto \frac{S_{sy}}{\langle B \rangle^{1+\alpha}},
\end{equation}
where $\alpha\simeq 1$ is the radiation spectral index.
The assumption that Coma radio halo is produced by secondary \epm,
as opposed to high energy \cre of different origin, 
allows us to also fix the total number of parent CR ions, 
$N_{cri}$, and the ensuing \pnd fluxes, 
$F_{\pi^0 \rightarrow \gamma\gamma}$. In fact,
since the steady-state total number of secondary CR electrons 
accords to 
\begin{equation} \label{norm0.eq}
 N_{e^\pm} \propto N_{cri} \langle \rho_{gas}\rangle  
\frac{1}{1+U_B/U_{\rm CMB}}
\end{equation}
based on this and eq. (\ref{norm1.eq}) one finds 
\begin{equation} \label{norm2.eq}
F_{\pi^0 \rightarrow \gamma\gamma} 
\propto N_{cri}\langle \rho_{gas} \rangle \propto S_{sy} 
\frac{1+\frac{U_B}{U_{\rm CMB}}}{\langle B \rangle ^{1+\alpha}},
\end{equation}
where $U_B= \langle B \rangle ^2 /8\pi$ and $U_{\rm CMB}$ 
are the energy density in magnetic field and CMB 
radiation field, respectively.
For the synchrotron flux needed in (\ref{norm1.eq})
and (\ref{norm2.eq}) we adopted the value $S_{1.4GHz} = 640$ mJy 
measured at 1.4 GHz by \citet{deissetal97}. 
When extracting synthetic spectra in the next section
we consider two cases for the magnetic field strength 
corresponding to $\langle B \rangle\sim 0.15\mu$G and 
$\langle B \rangle\sim 0.5\mu$G \citep{kkdl90}.
The former choice implies a ratio of CR ions to thermal energy
about 30 \% as opposed to almost 4\% for the latter. This second
case is assumed to produce the synthetic maps described below.
Notice that its ``renormalized'' values would imply a lower CR 
ion acceleration efficiency then described in \S \ref{injac.se}
and would also correspond to a parameter $R_{e/p} \sim 0.05$.

It is worth pointing out that according to eq. (\ref{norm1.eq}) 
and (\ref{norm2.eq}) smaller values of $\langle B \rangle$ 
than assumed here would imply correspondingly 
higher fluxes of \ic emission and \gr from \pnd. 
For very weak fields, then, most of the assumed radio 
emitting particles could not be of hadronic origin 
due to EGRET upper limits on \gr flux from Coma cluster
\citep[][\cf also fig. \ref{specsl.fig}]{blco99,mjkr01}.
In the opposite limit of stronger magnetic fields than assumed
here, less emitting particles would be required to produce the 
same radio emission. As a consequence, the associated \ic flux
is reduced proportionally to the inverse of the magnetic energy
density [\cf eq. (\ref{norm1.eq})].
However, as indicated by the expression in eq. (\ref{norm2.eq}),
for $\alpha \simeq 1$ and 
$\langle B \rangle \gg \sqrt{U_{\rm CMB}/8\pi}\sim 3.3~\mu$G,
the \gr production from \pnd reaches only a floor value.
This is because even if the higher magnetic field
implies a smaller number of radio emitting \epm particles,
in steady state a minimum rate of hadronic interactions is necessary
to compensate for the increased synchrotron losses. In fact 
the enhanced synchrotron emission rate from the total particle
distribution ($\propto \langle B \rangle ^{1+\alpha} \sim 
\langle B \rangle ^2$) is counterbalanced by the reduced number of 
steady state emitting \epm [$\propto N_{e^\pm}\simeq 
1/\langle B \rangle ^2$ - see eq. (\ref{norm0.eq})].
Thus in secondary models of radio halos, 
there is a lower limit to the expected \gr flux from \pnd. 

In fig. \ref{maps.fig} we present synthetic maps of the 
integrated photon flux above 100 keV (top panels) and 100 MeV 
(bottom panels) for an assumed $\langle B\rangle \simeq 0.5 \mu$G. 
Right panels are associated with \ic emission 
from primary \cre whereas left panels correspond to emission 
due to \ic from secondary \epm (top-left) and \pnd (bottom-left).
The top and bottom panels of fig. \ref{maps.fig} are combined
in the left and right panels of fig. \ref{maph.fig} respectively 
to produce a synthetic map of the total photon flux above 100 keV 
and 100 MeV. (For completeness these maps also include negligible 
contributions from the additional processes that were investigated 
in the previous section, \eg NTB.) A synthetic map associated with 
the (not rescaled) 
bolometric X-ray emission from thermal bremsstrahlung is also 
shown in the central panel of fig. \ref{maph.fig} to 
allow for comparison between thermal and non-thermal processes.
Non-thermal maps are presented in units 
of ``ph cm$^{-2}$ s$^{-1}$ arcmin$^{-2}$'' whereas thermal X-ray
surface brightness is in units ``erg  cm$^{-2}$ s$^{-1}$ arcmin$^{-2}$''. 

The synthetic images in fig. \ref{maps.fig} show that the 
non-thermal emissivity is remarkably 
extended. In accord with the previous section we find that
the emission from \pnd (bottom left) and \epm (top left) 
is confined to the cluster core. There it creates a diffuse 
halo which rapidly fades with distance from the center. Also,
as to be expected, there is a strong correlation between the
spatial distribution of both these emission components 
and that of the thermal X-ray emission (fig. \ref{maph.fig}).
On the other hand, \ic emission from primary
\cre is distributed over a much more extended area. Moreover, 
it is characterized by a strikingly rich and irregular 
structure. This is a direct reflection of the complex 
``web'' of shocks that reside at the outskirts of galaxy 
clusters first pointed out by \cite{minetal00}. 
The morphology of the emissivity in fig. \ref{maps.fig} and 
\ref{maph.fig} is quite similar, although noticeably the central 
diffuse emission stands out more prominently in the high energy
\gr map. 

\subsection{Synthetic Spectra} \label{synspe.se}

\begin{figure}
\includegraphics[width=0.5\textwidth ]{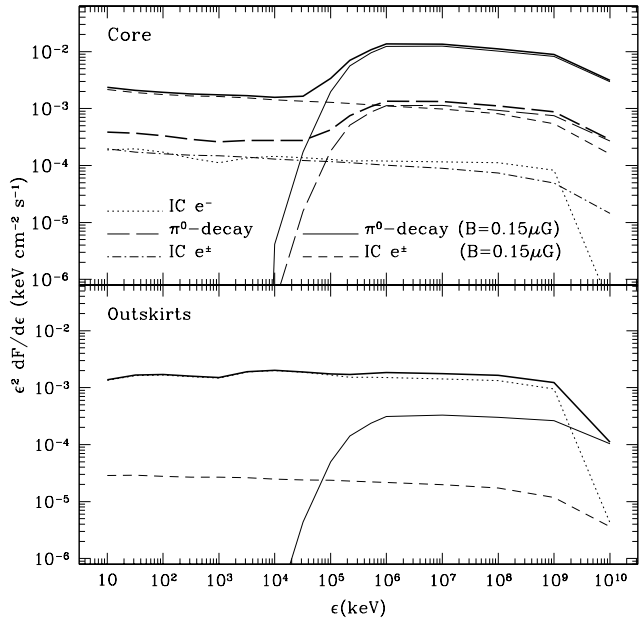}
\caption{\textit{Synthetic spectra extracted from the 
core region (top) and the outskirts region (bottom).}}
\label{specom.fig}
\end{figure}

In order to assess the separability of the 
different emission components, in fig. \ref{specom.fig} we have 
produced synthetic spectra taken from a core (top) and an outskirts 
region (bottom). The extension of the core region corresponds to 
an angular size of $1^o$ (or 2 Mpc diameter at the red-shift of 
Coma cluster) whereas the outskirts region is defined as an annular 
ring with inner and outer radii of $0.5^o$ and $1.5^o$ 
(or 1 Mpc and 3 Mpc at the red-shift of Coma cluster) respectively. 
For the core region we consider two values for the 
magnetic field strength (0.15 and 0.5 $\mu$G). 
For each emission region in fig. \ref{specom.fig}
we plot the predicted flux contribution due to the  
main emission processes considered so far namely, 
\ic emission from primary \cre (dot) and secondary \epm 
(dash and dot--dash), and \pnd (solid and long--dash). 
Notice that as the magnetic field strength is increased
from 0.15 to 0.5 $\mu$G the fluxes due to \ic from \epm and \pnd drop
by about a factor $\sim 10$ (\cf eq. \ref{norm1.eq} and \ref{norm2.eq}).
Fig. \ref{specom.fig} shows that at high photon energy ($>$ 100 MeV) 
the spectrum of radiation is indeed dominated 
by $\pi^0$-decay in the core region (top panel) 
and by \ic emission from primary \cre in the outer 
region (bottom).
In the latter case, the residual \pnd component is actually due 
to the presence in the field of view of a small structure 
north-west of the selected object (see fig. \ref{maps.fig}; 
since this is not part of the main cluster, 
we do not renormalize it according to various 
magnetic field values as it was done for the core emission).
In principle any contribution such as this can be removed by 
excision of emission regions associated with thermal X-rays 
from structures appearing in the field of view.
Below $\sim$ 100 MeV the flux from the 
outskirts region is still strongly dominated by \ic 
emission from primary \cre. The contribution from the latter
is much reduced in the narrower field of view of the core
region. However, for the larger field case  
it is still significant and at the level of \ic emission from 
secondary \epm. The relative amount of radiation flux 
from these two components further depends on 
the actual shock structure subtended by the field of view.
Nevertheless the point is that for magnetic field of a few 
tenths of $\mu$G they are expected of comparable intensity. 
In order to separate them out one could measure the radiation flux
as a function of radial distance in the outer regions, where it
presumably is only due to CR electrons accelerated at 
accretion shocks, and then extrapolate its value for the core region. 
For this purpose, an angular resolution $\sim 10^\prime$ or so should be 
sufficient.

Finally, above the spectra are characterized by a cutoff 
which appears at photon energies about 1 TeV. For \ic emission 
from primary e$^-$, for which this feature is sharpest,
this is mostly due to the maximum momentum, $p_{max}=20$ TeV,
of the accelerated particles.
For the other processes, it is rather caused by absorption due
to the reaction $\gamma\gamma \longrightarrow e^+e^-$.
In fact, pair production becomes important at TeV energies due 
to the presence of diffuse background radiation field at
optical/infrared wavelengths. The spectra in fig. \ref{specom.fig}
were thus corrected through a factor $\exp(-\tau_{\gamma\gamma})$
and by assuming an optical depth for pair production 
\citep[\eg][]{coah99}
\begin{equation}
\tau_{\gamma\gamma}(\varepsilon_\gamma) \simeq
0.14 \; \left( \frac{\varepsilon_\gamma}{\rm 1\; TeV} \right) \;
\left( \frac{u(\varepsilon_*)}{2\times 10^{-3} {\rm eV\; cm}^{-3}} \right) \;
\left( \frac{z_{Coma}}{0.023}\right) \; h^{-1}
\end{equation}
where $\varepsilon_*=4m_e^2c^4/\varepsilon_\gamma$ is the 
target photon energy and $u(\varepsilon_*)$ is the energy 
density carried by the background radiation taken from 
\citet{dwar98}.

\section{Discussion} \label{disc.se}
\begin{figure}
\includegraphics[width=0.50\textwidth ]{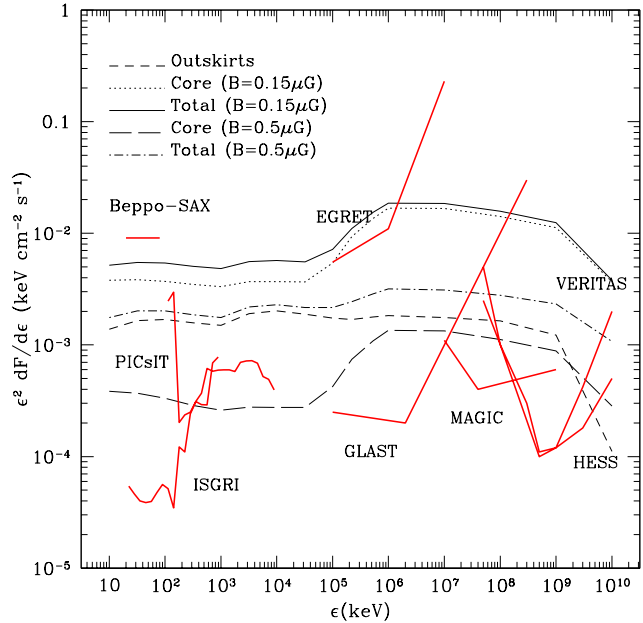}
\caption{\textit{Total radiation spectra extracted
from the same spatial regions illustrated in fig. 
\ref{specom.fig} 
-- core for two different magnetic field values
(dot and long--dash), outskirts (short dash) and 
their summation (solid and dot--dash) -- 
compared to nominal sensitivity limits 
of future \gr observatories (thick-solid lines). 
For INTEGRAL-IBIS imagers (ISGRI and PICsIT) the curves 
correspond to a detection significance of 3$\sigma$ 
with an observing time of $10^6$ s. All other sensitivity 
plots refer to a 5$\sigma$ significance. EGRET and GLAST 
sensitivities are shown for one year 
of all sky survey whereas for Cherenkov telescopes
(MAGIC, VERITAS and HESS) for 50 hour exposure
on a single source. }}
\label{specsl.fig}
\end{figure}

In fig. \ref{specsl.fig} we plot sensitivity limits of 
planned \gr observatories together with
the total radiation spectra from the core (dotted line) 
and outskirts (dashed line) regions that were presented in 
fig. \ref{specom.fig}. The sensitivity limits are plotted for
point sources. For extended sources it will be worse by roughly
a factor that goes as $\theta_{source}/ \theta_{inst}$ that is
the ratio between the source size and the angular resolution
of the instrument. 
For Cherenkov telescopes $\theta_{inst} \sim 0.1^o $
In any case, 
according to this plot, several future experiments should be 
sensitive enough to detect the computed non-thermal emission 
at most photon energies.
In particular the IBIS imager onboard INTEGRAL should readily 
measure the flux between 100 keV and several MeV. 
In addition GLAST and Cherenkov telescopes
should be able to detect both the core \gr emission 
from \pnd as well as the \ic flux directly
produced at accretion shocks above 100 MeV and 10 GeV respectively.
One should notice that resolving sharp and complex structures 
such as those reported in fig. \ref{maps.fig} and \ref{maph.fig} is a real
challenge, especially for coded mask techniques employed by several
current and planned \gr imagers. Although a nominal resolution of $\sim
10^\prime $ in principle should be sufficient to identify some of the 
bright shock features, the final outcome of such measurements 
will hinge on the actual source fluxes and instruments performance.

The measurement of the non-thermal radiation spectra at 
several photon energies spanning the range illustrated 
in fig. \ref{specom.fig} provides important
information about the physical conditions in clusters.
First, besides the mere detection of HXR 
and \gr radiation from clusters, important information is 
embedded in the spatial distribution of their surface brightness.
In particular the extended emission component corresponds to 
the location of accretion shocks. Merger shocks have occasionally 
been observed in the core of clusters as relatively weak 
temperature jumps. However, strong accretion shocks have yet 
to be observed {\it directly}. Thus, detection and imaging
of \ic emission from primary electrons would provide an opportunity 
to directly observe these accretion shocks given their wide angular 
extension and their unique morphology.

In addition, the flux about 100 keV will give us the first direct 
estimate of the energy density of CR electrons and, together 
with radio measurements, will allow an estimate of 
(or upper limits on) the ICM magnetic field. 
In this respect for a correct interpretation of the data 
it will be necessary to account properly for the fluxes 
contributed by both the CR electrons in the 
core and those directly accelerated at the outlining shocks.
In fact, the former likely propagate in relatively highly
magnetized regions and produce a substantial radio emission.
On the other hand, because of the expected decline of the
magnetic field strength toward the outer regions, the latter 
might generate only a weak radio emission but, nonetheless, 
a strong \ic flux. Since, as shown in fig. \ref{specom.fig}, 
the HXR flux produced by this second component can easily 
dominate the total flux at this spectral range 
(the details will depend on the assumed 
normalizations on acceleration efficiency and average 
magnetic field strength) separating the contributions from 
CRs in the cluster core and external accretion shocks will be 
an important step for correctly measuring ICM magnetic fields.
A similar point was qualitatively discussed already in 
\citet{mjkr01} and was also addressed in \citet{bsfgf01a} 
in the context of a their ``multiphase'' acceleration 
model. In particular \citet{bsfgf01a} pointed out that the
magnetic field in the outer region, where most of the 
\ic emission is produced due to the larger emitting volume,
can be much lower than in the core 
where the radio emission originates. In that model all the radio
and HXR emitting particles are accelerated with the same 
mechanism, whereas in our
model the CR electrons that generate the HXR flux are accelerated
in the outer accretion shocks and never make it to the cluster core. 

It is worth mentioning that our estimated non-thermal 
flux from \ic emission above 20 keV is quite smaller 
than the recently reported measurements of excess of HXR 
radiation with respect to thermal emission. 
For the case of Coma cluster our predictions, as illustrated in fig. 
\ref{specsl.fig}, fall short by a factor of several 
\citep{fufeetal99,regrbl99}. An analogous estimate,
obtained for A2256 after appropriate rescaling, indicates 
a similar discrepancy by a factor $\sim 30$ \citep{fufeetal00}, 
although the upper limits on A3667 are respected by our predictions
\citep{fufeetal01}. The above 
discrepancies between prediction and reported detections 
could be readily improved by assuming that the electron acceleration 
efficiency at shocks is larger by an order of magnitude with 
respect to what assumed here \cite[see also][]{lowa00}. 
Interestingly, since these electrons are located 
at the cluster outskirts and need not be responsible for the radio 
halo emission, their \hxr emission would not constraint the 
ICM magnetic field detected through Faraday rotation measures. 
However, as it is apparent from fig. \ref{specsl.fig},
this is at odds with EGRET experimental upper limits on 
the \gr flux above 100 MeV \citep{sreeku96,reimer99,reposrma03},
which allow for the 
acceleration efficiency adopted here to be increased by at most
a factor $\sim 2$ \citep{min02b}. This constraint, however, 
holds true only as long as: (1) CR \cre are accelerated above momenta 
$\sim 100$ GeV/c and (2) the spectrum of the accelerated particles 
is not much steeper than computed here. These conditions, though,
seem to be easily fulfilled for the case of accretion shocks.
Thus, in this respect, 
if all of the HXR excess emission is diffuse in nature and not
associated with individual sources, turbulent and/ or second order
Fermi acceleration models may provide a more natural explanation
because a high momentum cutoff in this range arises more naturally 
in those models (provided of course enough acceleration efficiency).

One of the most awaited experiments is related to the measurement 
of the \gr flux at and above 100 MeV. This is important in order to 
convalidate or rule out secondary \epm models for radio 
halos in galaxy clusters 
\citep{dennison80,vestrand82,doen00,blco99,mjkr01}
and in order to estimate the non-thermal CR pressure 
there \citep{mrkj01}. In this perspective the above authors 
estimated for nearby clusters 
the \gr flux produced from the decay of neutral 
pions produced in p-p inelastic collisions. As pointed out in 
\citet{min02b}, however, 
\gr from \ic emission can also be substantial and at the 
same level as that from \pnd. Therefore, once again 
for a correct interpretation of the measurements, 
separation of these two components will be required.
Estimating the \gr flux from \ic emission due to shock accelerated
CR electrons will be instrumental for addressing another
issue of great interest. That is, the contribution of this 
emission mechanism to the cosmic \gr background 
\citep{lowa00,min02b,keshetetal02},
for which so far the only constraint is provided by an upper 
limit based on EGRET experiments \citep{min02b}.

Photon energies of order $\sim 10-100$ GeV to $1-10$ TeV, 
will be investigated by Cherenkov telescopes. 
Such measurements will be complementary to that carried out
at lower energies. 
Because the radiated energy 
spectrum is directly connected to the distribution function 
of the emitting particles, measuring the flux at different 
energies will provide information about the acceleration
mechanism. In particular, the observed spectrum could 
differ from our predictions if the accretion shocks were 
modified by CR pressure. That in fact would cause the
radiation spectrum to soften at low energies and to become
harder toward higher energies, up to an energy cut-off 
\cite[\eg][]{madr01,beel99,kajo02}.
Given the different environmental 
conditions with respect to supernova remnants, these 
are interesting issues for a broad exploration of 
shock acceleration physics. 
Independent of the shock dynamics, measurements in this 
photon energy range should help determining the maximum 
momentum of the accelerated CR electrons and, perhaps, even of 
CR ions in case these maxima produce spectral cutoff before 
attenuation by $\gamma\gamma$ absorption becomes important. 
For the electronic component this is not excluded and 
could allow a clearer specification of the energy range 
in which they can contribute to the cosmic \gr background.

\section{Summary \& Conclusions} \label{summ.se}

We have explored the spatial and spectral properties of
non-thermal emission at \gr energies. For the purpose we
carried out a simulation of structure formation including
the evolution of baryonic gas, dark matter, 
CR ions and electrons accelerated at cosmic shocks as well
as secondary \epm generated in inelastic p-p collisions. 
We estimated the radiation flux between 10 keV and 10 TeV
from CRs in collapsed structures due to \ic emission, 
\pnd and NTB and made specific predictions 
for the case of Coma cluster.
We have assessed the importance of distinguishing among the
contribution from different CR components for a 
correct interpretation of future experimental results 
and we have outlined a strategy to do so. 
Our conclusions are summarized as follows:

\begin{itemize}

\item Two main regions for production of non-thermal radiation 
in clusters/groups of galaxies were identified: the core also 
bright in thermal X-ray 
and the outskirts region where accretion shocks occur.

\item The chief radiation mechanism at all \gr energies
in the outskirts region is \ic emission from shock accelerated 
CR electrons, provided that a fraction of a percent of the shock 
ram pressure is converted into CR electrons. A clear detection 
of this component and of its spatial distribution will allow us
direct probing of cosmic accretion shocks.
Such evidence would corroborate recent findings about 
extra-galactic radio emission from large scale shocks
\citep{bagchietal02}.

\item In the cluster core, \gr emission above 100 MeV is dominated
by \pnd mechanism. At lower energies, \ic emission from 
secondary \epm takes over. However, \ic emission from 
shock accelerated electrons projected onto the cluster core
will not be negligible in general.

\item Separating the aforementioned emission components is 
important for a correct interpretation of the experimental data.
This can be achieved in principle by measuring the spatial 
distribution of the detected emission.

\item Measuring the non-thermal spectrum will provide us with 
knowledge regarding: the injection efficiency as well as the 
ram pressure to CR pressure conversion efficiency for both 
electrons and ions; the energy range in which \ic emission from
CR electrons accelerated at accretion shocks can contribute
to the CGB; the CR content in galaxy clusters; the dynamical
conditions (CR modified or not) of the accretion shocks.

\end{itemize}

\section*{Acknowledgments}
This work was carried out at the Max-Planck-Institut 
f\"ur Astrophysik under the auspices of the European 
Commission for the `Physics of the Intergalactic Medium'.
I am grateful to E. Churazov, T. En{\ss}lin, M. Gilfanov,
F. Aharonian and I. Susumu for useful discussion.
The computational work was carried out at the 
the Rechenzentrum in Garching operated by the 
Institut f\"ur Plasma Physics and the Max-Planck Gesellschaft.

\bibliographystyle{apj}
\bibliography{papers,books,proceed}

 \label{lastpage}
\end{document}